\documentclass[adp,fleqn]{w-art}
\usepackage{times}
\usepackage{w-thm}
\usepackage{graphicx}

 \newcommand{\beq}{\begin{equation}}
 \newcommand{\eeq}{\end{equation}}
 \newcommand{\boma}[1]{\mbox{\boldmath$#1$\unboldmath}}

 \newcommand{\bpr}{\dot{\boma{r}}}
 \newcommand{\hsp}{\hspace{5mm}}
 \newcommand{\calo}{{\cal O}}

 \newcommand{\bA}{\boma{A}}
 \newcommand{\bB}{\boma{B}}
 \newcommand{\bE}{\boma{E}}
 \newcommand{\bj}{\boma{j}}
 \newcommand{\bJ}{\boma{J}}
 \newcommand{\bl}{\boma{\ell}}
 \newcommand{\bL}{\boma{L}}
 \newcommand{\bp}{\boma{p}}
 \newcommand{\bP}{\boma{P}}
 \newcommand{\br}{\boma{r}}

 \newcommand{\kom}{\omega}
 \newcommand{\gom}{\Omega}
 \newcommand{\calm}{\cal M}
 \newcommand{\ppsik}{\dot{\psi}_{\mbox{\tiny K}}}
 \newcommand{\Tk}{T_{\mbox{\tiny K}}}
 \newcommand{\rerde}{R_{\mbox{\tiny E}}}

 \newcommand{\muQ}{\mu_{\mbox{\tiny Q}}}
 \newcommand{\bmuq}{\boma{\mu}_{\mbox{\tiny q}}}
 \newcommand{\bmuQ}{\boma{\mu}_{\mbox{\tiny Q}}}
 \newcommand{\komq}{\kom_{\mbox{\tiny q}}}
 \newcommand{\bkomQ}{\boma{\kom}_{\mbox{\tiny Q}}}
 \newcommand{\bgomq}{\boma{\gom}_{\mbox{\tiny q}}}
 \newcommand{\qe}{q_{\mbox{\tiny E}}}
 \newcommand{\qb}{q_{\mbox{\tiny B}}}
 \newcommand{\Qe}{Q_{\mbox{\tiny E}}}
 \newcommand{\Qb}{Q_{\mbox{\tiny B}}}
 \newcommand{\ppsi}{\dot{\psi}}
 \newcommand{\LT}{{\mbox{\tiny LT}}}
\begin{document}
\DOIsuffix{theDOIsuffix}
\Volume{51}
\Issue{1}
\Month{01}
\Year{2003}
\pagespan{1}{}



\title[The gravitomagnetic clock effect]{The gravitomagnetic clock effect and its possible observation}


\author[H.\ Lichtenegger]{H. Lichtenegger\inst{1}%
  \footnote{Corresponding author \quad E-mail: {\sf herbert.lichtenegger@oeaw.ac.at},
            Phone: +43\,316\,4120\,554, Fax: +43\,316\,4120\,590}}
\address[\inst{1}]{Institut f\"ur Weltraumforschung,
        \"Osterreichische Akademie der Wissenschaften, A-8042 Graz, Austria}
\author[L. Iorio]{L. Iorio\inst{2}}
\address[\inst{2}]{Viale Unit$\grave{\rm a}$ di Italia 68, 70125, Bari, Italy}
\author[B.\ Mashhoon]{B. Mashhoon\inst{3}}
\address[\inst{3}]{Department of Physics and Astronomy, University of Missouri-Columbia,
        Columbia, Missouri 65211, USA}
\begin{abstract}
The general relativistic gravitomagnetic clock effect involves a
coupling between the orbital motion of a test particle and the
rotation of the central mass and results in a difference in the
proper periods of two counter--revolving satellites. It is shown
that at ${\cal O} (c^{-2})$ this effect has a simple analogue in
the electromagnetic case. Moreover, in view of a possible
measurement of the clock effect in the gravitational field of the
Earth, we investigate the influence of some classical perturbing
forces of the terrestrial space environment on the orbital motion
of test bodies along opposite trajectories.
\end{abstract}
\maketitle                   






\section{Introduction}
\label{sect1}
According to Einstein's relativistic theory of gravitation, the rotation of a mass induces a
particular temporal structure in its surroundings that is clearly revealed by the gravitomagnetic
clock effect [1--4]. The purpose of this paper is to discuss this basic effect and emphasize its
close correspondence with an analogous electromagnetic phenomenon, thereby extending our previous
work on an alternative derivation of the gravitomagnetic clock effect \cite{Iorio_et_al:2002}. Moreover,
we discuss the prospects for the observation of this interesting effect.

Gravitomagnetism is expected to be ubiquitous, yet its detection is a very difficult venture.
Concerning the terrestrial space environment, the GP-B mission \cite{Everitt_et_al:2001},
aimed at the detection of the gravitomagnetic precessions of the spins \cite{Schiff:1960} of four gyros
carried aboard the spacecraft, has been successfully launched in April 2004 and the results are expected
to be announced sometime in 2007 \cite{www:stanford}; the anticipated accuracy is $1\%$ or better. A test
of another gravitomagnetic effect involving the Lense-Thirring precessions [9--10] of the orbits of the
LAGEOS and LAGEOS II Earth satellites has been carried out recently [11--12]. The claimed accuracy is $5\%$
at one sigma; for a critical discussion of such a test see \cite{Iorio:2006}. The values of the Lense-Thirring
precessions of the inner planets of the Solar System have recently been found to be in general agreement
\cite{Iorio:2005} with the latest determinations of the planetary perihelia advances \cite{Pitjeva:2005},
although the errors are still large. However, they might be measured with a better accuracy in the near
future when more data will be gathered and processed.


Among the various gravitomagnetic manifestations which might become observable in the solar system
within the next few years, the gravitomagnetic clock effect, which involves, among other things,
the difference of the sidereal periods of two free counter-orbiting particles about a rotating
massive body, is of special interest since it contradicts the common notion of the ``dragging of
inertial frames". In the following we will show that this effect has a simple correspondence in
the electromagnetic case by considering the motion of a test charge in a dipolar magnetic field,
thereby making the nature of the clock effect more transparent (see also \cite{Iorio_et_al:2002}).
Finally, the paper ends with a brief discussion of the required accuracy in the determination
of the orbital elements in order to test the clock effect in a possible experiment with two
counter-revolving satellites about the Earth.

\section{The electromagnetic scenario}
\label{sect2}
We consider a spherical body of homogeneously distributed mass $M$ with a charge $Q$,
which rotates with a small angular velocity $\bkomQ$, thereby producing a weak magnetic
dipole moment $\bmuQ$. The axis of rotation and the direction of $\bmuQ$ shall coincide
with the positive $z$-axis of the background inertial coordinate system. We study the
slow motion of a test charge of mass $m\ll M$ and charge $q\,(qQ<0)$ about the central body such
that in the absence of the perturbation due to $\bmuQ$, the motion takes place along a
circular inclined orbit, where the nodal line coincides with the $x$-axis, as shown in
Fig. 1. Once the magnetic perturbation is "turned on", the orbit will no longer be planar;
however, the orbital radius can remain constant to first order in the perturbation [5].
Throughout this paper it will be assumed that the test particle moves along a {\it spherical}
orbit of constant radius and that the magnetic field is weak so that second-order effects
may be neglected. Moreover, we will ignore all radiative and special relativistic effects
on the motion of the electric charges.

In the following we show that the motion of the test mass is modified due to the weak
dipolar magnetic field in two ways: (a) the orbital plane precesses about the spin axis
of the central body, and (b) the along-track velocity of the test particle is changed.

\begin{vchfigure}[tb]
\includegraphics[width=.5\textwidth]{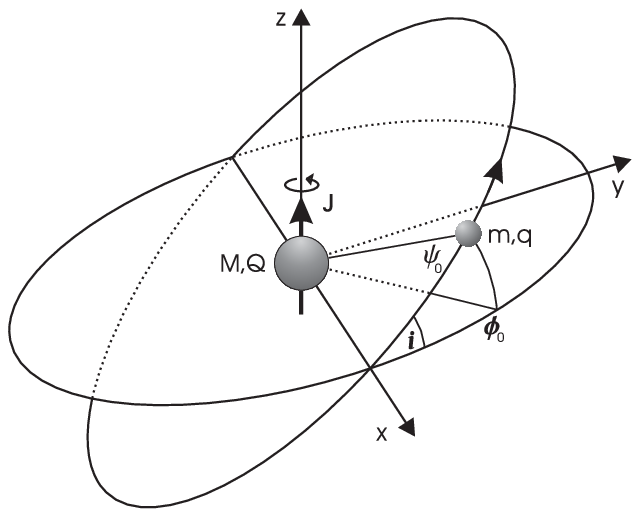}
\vchcaption{A test charge ($m,\,q$) moves about a central slowly rotating body ($M,\,Q$).
The figure depicts the initial configuration when the rotation of the source is ``turned on".
The unperturbed orbit of the test mass is assumed to be circular; moreover, $\psi_0$ is the
initial angle in the orbital plane, while $\phi_0$ is the corresponding angle measured along
the equatorial plane to which the orbital plane is tilted by the inclination angle $i$.}
\label{fig:1}
\end{vchfigure}

\subsection{The equations of motion}
The rigid rotation of the central charge produces a current
density $\bj$ which involves a magnetic dipole moment
\beq
\bmuQ=\frac{1}{2c}\int\br\times\bj\,d^3 r
\eeq
and hence via the magnetic vector potential
\beq\label{elektro.vekpot-e}
\bA=\frac{\bmuQ\times\br}{r^3},
\eeq
the magnetic dipole field
\beq\label{elektro.dipol-e}
\bB=\boma{\nabla}\times\bA=\frac{3\br(\br\cdot\bmuQ)-\bmuQ r^2}{r^5}.
\eeq
The motion of the test mass in the electric and magnetic field produced by the
central charge is governed by the Lorentz force,
\beq
\boma{F}=q\left(\boma{E}+\frac{1}{c}\,\dot{\br}\times\bB\right),
\eeq
where $\br$ is the position vector joining $Q$ with $q$, an overdot denotes
differentiation with respect to time, $\bE=(Q/r^3)\br$ is the electric
field and $\bB$ is given by (\ref{elektro.dipol-e}). For later purposes, we
formally introduce electric and magnetic charges $\Qe,\,\qe$ and $\Qb,\,\qb$,
respectively, and write the equations of motion as
\beq\label{elektro.bewgl}
\ddot{\br}=\frac{\Qe\qe}{m}\frac{\br}{r^3}+\frac{\qb}{mc}\,\dot{\br}\times\bB,
\eeq
with $Q=\Qe=\Qb$ and $q=\qe=\qb$. It must be emphasized that the magnetic charge
(such as $\Qb$ or $\qb$) introduced here has no connection with the magnetic monopole
strength that is assumed to be zero in this work.

\subsection{Precession of the orbital plane}
Since the orbiting test charge possesses a dipole moment
\beq
\bmuq=\frac{\qb}{2mc}\bl,
\eeq
where $\bl$ is the orbital angular momentum, a torque $\boma{M}=\bmuq\times\bB$
will act upon $\bmuq$, leading to a precession of the orbital plane about the $z$-axis
with angular velocity $\bgomq$ according to $\dot{\bl}=\bgomq\times\bl$.

In order to find $\bgomq$, we first determine the vector potential $\bA$. Since the
magnetic field $\bB$ is associated with the uniform rotation of the central charge of
angular momentum $\bJ=I\bkomQ$, where $I$ is the moment of inertia of $M$, the dipole
moment of the central spherical object becomes
\beq\label{elektro.gyro}
\bmuQ=\frac{\Qb}{2Mc}\bJ=\frac{\Qb I}{2Mc}\bkomQ,
\eeq
or, upon inserting into (\ref{elektro.vekpot-e})
\beq
\bA=\alpha\,\bkomQ\times\br,
\eeq
with $\alpha=\Qb I/(2Mcr^3)$. Vector multiplication of (\ref{elektro.bewgl}) with $m\br$
and using $\bB=\alpha\boma\nabla\times(\bkomQ\times\br)$ as well as substituting
$\bp=m\dot{\br}$ in the canonical momentum $\bP=\bp+(\qb/c)\bA$ yield
\beq
\dot{\bL}-\frac{\qb}{c}\left(\dot{\br}\times\bA+\br\times\dot{\bA}\right)=
 \frac{2\qb\alpha}{c}\br\times(\dot{\br}\times\bkomQ),
\eeq
where $\bL=\br\times\bP$. Finally, since $\bkomQ$ and $r$ do not depend upon time, we arrive at
\beq
\dot{\bL}= -\frac{\qb\alpha}{c}\bkomQ\times(\br\times\dot{\br})=
 \bgomq\times\bL+\frac{\Qb\qb^2}{2Mmc^3r^3}\bJ\times(\br\times\bA),
\eeq
where
\beq\label{elektro.pre}
\bgomq=-\frac{\qb\alpha}{mc}\bkomQ=-\frac{\Qb\qb}{Mm}\frac{\bJ}{2c^2r^3}.
\eeq
The last term in equation (10) is of $\calo(c^{-4})$, since $\bA = \calo(c^{-1})$, and is
second order in the magnetic perturbation; therefore, it is negligible in accordance
with our approximation scheme. It follows that the canonical angular momentum precesses
with frequency $\bgomq$ that is ``slow" in comparison with the ``fast" orbital motion.

We note that the product $\Qb\qb$ will always be negative, since we are considering
attractive forces only, therefore $\bgomq$ and $\bJ$ point in the same direction. It
is interesting to note that in our approximation scheme, the orbital speed as well as the
magnitude of the orbital angular momentum remains constant in accordance with equation (5).
What we have demonstrated here for canonical angular momentum also holds true, when averaged
with respect to the ``fast" orbital motion, for $\bl=\br\times\bp$. That is, it can be shown
that on the average, the orbital plane of the test charge is ``dragged" in the same sense as
the rotation of the central charge.

\subsection{The change in the orbital velocity}
In the following we will calculate the change in the along-track
velocity due to the magnetic dipole field (\ref{elektro.dipol-e}).
The result can be understood heuristically by determining the
magnetic flux through the orbital plane. If we assume that the
magnetic field is switched on and increases from zero to
$\boma{B}$, then by means of Faraday's law we can calculate the
electric field strength $E$ induced in the orbital plane due to
the varying magnetic field and obtain
\beq\label{elektro.e-psi}
E=\frac{1}{2cr^2}\frac{d\muQ}{d t}\cos i,
\eeq
where $i$ is the inclination angle. From (\ref{elektro.e-psi}) and
$\qb E=m{dv}/{dt}=mr{d\dot\psi}/{dt}$, the change in the angular velocity
$\Delta\ppsi$ caused by the dipolar magnetic field becomes
\beq\label{elektro.deltapsip}
\Delta\ppsi=\frac{\Qb\qb}{Mm}\frac{J}{4c^2r^3}\cos i;
\eeq
this result can be obtained directly starting from equation (5). Since $\Qb\qb<0$,
we see that the test charge moves slower in the prograde direction (i.e. in the
direction of the rotation of the source), and faster in the retrograde direction
as compared to a situation without the magnetic field. This behavior is also
intuitively confirmed by the subsequent evolution of the motion under the action
of the Lorentz force: for the prograde motion the Lorentz force points away from
the source and weakens the attractive Coulomb force. We note that this feature is
independent of the sign of the revolving charge, because the exchange of the signs
of the central and orbiting charge will also change the direction of the magnetic
field.

The change in the velocity of the revolving charge can also be
anticipated by virtue of Lenz's rule: the induced current will
have such a direction as to counteract its own source. Indeed,
let us replace the rotating source by a single current loop and
the orbit of the revolving charge by a conducting wire concentric
to the inner loop. Any change $\Delta I$ of the inner current will
induce a current in the outer wire having the same direction as
$\Delta I$, thus leading to an increase of the magnetic flux
through the inner loop. This additional flux in turn induces a
current opposite to $\Delta I$ and thus opposes a change of the
inner current.

\subsection{Time of revolution}
According to (\ref{elektro.deltapsip}), the velocity of the test
charge depends on the direction of the motion and thus its period
will be different for pro- and retrograde orbits. If the magnetic
field is absent, the Kepler period is $\Tk=2\pi/\ppsik$, with the
frequency $\ppsik=(|{\Qe\qe}|/mr^3)^{1/2}$, while in the presence
of a magnetic field, $\ppsi=\ppsik+\Delta\ppsi$ and the time of
revolution $T^*$ becomes
\beq\label{elektro.periode1}
T^*=\frac{2\pi}{\ppsi}\simeq\frac{2\pi}{\ppsik}\left(1-\frac{\Delta\ppsi}{\ppsik}\right)=
 \frac{2\pi}{\ppsik}\left(1-\frac{1}{\ppsik}\frac{\qb\muQ}{2mcr^3}\cos i\right).
\eeq
The difference in the period for two orbits with different inclinations
$i_1$ and $i_2$ is immediately found from (\ref{elektro.periode1}) and
(\ref{elektro.gyro}) to be
\beq
T_1^*-T_2^*=\frac{1}{2}\pi\frac{J}{Mc^2}(\cos i_1-\cos i_2)
\eeq
for electric charges. Note that this time difference, which refers to the closure
of the angle $\psi$ in the orbital plane, depends only on the specific angular
momentum of the source, but neither on the charge nor the distance $r$ of the test
mass, and it is maximum for opposite equatorial orbits (i.e. $i_1=0,\,i_2=\pi$ or
vice versa).

\subsection{The magnetic clock effect}
As demonstrated in the preceding paragraphs, the motion of a test charge
in a central Coulomb field together with a weak dipolar magnetic field
results in a slow precession of the orbital plane as well as a slightly
increased or decreased along-track velocity.

An inertial observer will define the period of the particle as the
time elapsed between two successive crossings of a fixed plane in
space, and consequently this period will differ from
(\ref{elektro.periode1}) due to the nodal precession of the
orbital plane. The time of revolution as measured by the inertial
observer can be calculated by noting that the particle's position
angle $\psi$ in the orbital plane is related to the corresponding
angle $\phi$ measured from the lines of the ascending nodes in the
equatorial plane by
\beq\label{elektro.kgl}
\tan\phi=\tan\psi\cos i.
\eeq
Now, let $\psi_0=\psi(t=0)$ and $\phi_0=\phi(t=0)$ denote
the initial position of the particle and $T$ be its period with
respect to the inertial frame, where $T=T^*+\Delta T$ and
$\Delta T\ll T,\,T^*$. The angle swept over by the particle after
completion of one revolution as seen by the inertial observer in the
orbital plane becomes $\psi_0+\ppsi T-\komq T\cos i$, while in the
equatorial plane this corresponds to $\phi_0+2\pi-\komq T$ and therefore
\beq
\tan(\phi_0-\komq T)=\tan(\psi_0+\ppsi T-\komq T\cos i)\cos i.
\eeq
By means of trigonometric relations and by noting that $\tan\ppsi T\simeq\ppsi\Delta T$,
$\tan\komq T\simeq\komq T^*$ and $\ppsi^{-1}\simeq\ppsik^{-1}(1-\Delta\ppsi/\ppsik)$ together
with (\ref{elektro.pre}) and (\ref{elektro.deltapsip}), we finally obtain to first order by
virtue of (\ref{elektro.kgl})
\beq\label{elektro.periode2}
T=\frac{2\pi}{\ppsik}\left[1-\frac{1}{\ppsik}\frac{\muQ\qb}{2mcr^3}
 \left(1-2\tan^2 i\cos^2\psi_0\right)\cos i\right].
\eeq
From this equation, the difference in the periods for two orbits with inclination $i_1$ and $i_2$
is found to be
\beq\label{elektro.tdiff}
T_1-T_2=\frac{\pi\qb}{|\Qe\qe|}\frac{\muQ}{c}\left[2\cos^2\psi_0
 \left(\tan^2 i_1\cos i_1-\tan^2 i_2\cos i_2\right)-\cos i_1+\cos i_2\right],
\eeq
which reduces for $\qe=\qb$ and $\Qe=\Qb$ to the magnetic clock effect
\beq\label{elektro.mce}
T_1-T_2=-\frac{1}{2}\pi\frac{J}{Mc^2}\left[2\cos^2\psi_0
 \left(\tan^2 i_1\cos i_1-\tan^2 i_2\cos i_2\right)-\cos i_1+\cos i_2\right];
\eeq
for opposite or supplementary orbits ($i_1=i,\, i_2=\pi-i$), this relation simplifies to
\beq\label{elektro.mce-opp}
T_1-T_2=\pi\frac{J}{Mc^2}\left(1-2\tan^2 i\cos^2\psi_0\right)\cos i.
\eeq
It should be emphasized that the dependence of the period (and of the time difference in turn)
on $\psi_0$ is due to the non-linear relation (\ref{elektro.kgl}).

It is interesting to note that the period for a prograde orbit is
longer than the Kepler period for low inclinations (remember that
$\qb\muQ<0$) and shorter for high inclinations, because in the
first case the lower orbital velocity implies $T>\Tk$, while in
the latter case the nodal precession, which makes the period to
decrease, dominates.

\section{The gravitomagnetic scenario}
\label{sect3}
In the case of slow motion and weak gravitational fields, the spacetime metric can be approximated by
\beq\label{elektro.gem-metrik}
g_{\mu\nu}=\eta_{\mu\nu}+h_{\mu\nu},
\eeq
where $\eta_{\mu\nu}={\rm diag}(-1,+1,+1,+1)$ is the Minkowski metric and $|h_{\mu\nu}|\ll 1$ is a small
correction so that only terms linear in $h_{\mu\nu}$ will be of significance.

\subsection{The equations of motion}
As in the electromagnetic case, the mass current density
$\bj_g$ associated with the rotation of the central mass with
angular momentum $\bJ$ produces a dipole moment
\beq
\boma{\mu}_g=\frac{1}{2c}\int \br\times\bj_g d^3 r
\equiv\frac{G}{c}\bJ
\eeq
with the corresponding vector potential
\beq 
\boma{A}_g=\frac{G}{c}\frac{\boma{J}\times\br}{r^3},
\eeq
which implies the gravitomagnetic dipole field
\beq
\bB_g=\boma{\nabla}\times\bA_g=\frac{G}{c}\frac{3\br(\br\cdot\bJ)-\bJ r^2}{r^5}
\eeq
analogous to (\ref{elektro.dipol-e}).

In the case of weak perturbations (\ref{elektro.gem-metrik}) and upon restricting to
static fields, the geodesic equation can be written in the form \cite{Mashhoon_et_al:2001b}
\beq\label{elektro.gem-bewgl1}
\boma{F}=-m\left(\bE_g+\boma{\cal{E}}_g+\frac{2}{c}\,\bpr\times\bB_g\right),
\eeq
where $\bE_g=GM\br/r^3$ and $\boma{\cal{E}}_g=\calo(c^{-2})$ is the post-Newtonian
gravitoelectric field. 
As has been first pointed out in \cite{Mashhoon:1993}, a complete correspondence with
electromagnetism can be achieved by introducing the notion of gravitoelectric and
gravitomagnetic charges, $\qe$ and $\qb$, respectively, and ignoring $\boma{\cal{E}}_g$.
The equation of motion (\ref{elektro.gem-bewgl1}) thus becomes
\beq\label{elektro.gem-bewgl2}
\ddot{\br}=\frac{GM\qe}{m}\frac{\br}{r^3}+\frac{\qb}{mc}\,\dot{\br}\times\bB_g,
\eeq
in perfect agreement with (\ref{elektro.bewgl}). The appearance of the
factor 2 in the ``magnetic" force term of (\ref{elektro.gem-bewgl1})
suggests $\qb=-2m$ as well as $\qe=-m$, where the additional minus sign
accounts for the fact that gravitational charges of opposite polarity,
like in the electromagnetic case, should be attractive. From
(\ref{elektro.gem-bewgl2}) we also read $\Qe=GM$ and consequently $\Qb=2GM$.

\subsection{The gravitomagnetic clock effect}
Because the equations of motion of electrodynamics and linear general relativity are formally equivalent,
in the weak-field and low-velocity limit gravitational phenomena are expected to be similar to those known
in electrodynamics. Therefore, by means of the substitutions derived in the preceding section,
\beq
\begin{array}{ccccccc}
\qe&\rightarrow& -m,&\hsp& \Qe&\rightarrow& GM,\\
\qb&\rightarrow& -2m,&\hsp& \Qb&\rightarrow& 2GM,
\end{array}
\eeq
we can directly translate electromagnetic phenomena into the corresponding gravitoelectromagnetic phenomena.
It is important to note that the ratio of the gravitomagnetic charge to the gravitoelectric charge is
always 2, which corresponds to the spin-2 character of linear gravitation. This is in agreement with
the fact that the ratio of the magnetic charge to the electric charge is always unity in Maxwell's spin-1
electrodynamics.

With $\muQ\rightarrow GJ/c$ from (\ref{elektro.gyro}) we immediately obtain by virtue of
(\ref{elektro.pre})
\beq
\Omega_\LT=\frac{2GJ}{c^2r^3},
\eeq
which is the well-known Lense-Thirring precession of the node of the orbital plane of a satellite.
Further, let $T_\pm$ denote the period of revolution in the same (+) and opposite (-) sense as the
rotation of the source, then (\ref{elektro.periode2}) translates into
\beq\label{elektro.gce}
T_\pm=\Tk\pm 2\pi\frac{J}{Mc^2}\left(1-2\tan^2 i\cos^2\psi_0\right)\cos i,
\eeq
which is equivalent to equation (35) of \cite{Mashhoon_et_al:1999} if terms of $\calo (c^{-4})$ are
neglected. Finally, for opposite equatorial orbits it follows from (\ref{elektro.gce}) that
\beq\label{elektro.gem-clock}
T_+-T_-=4\pi\frac{J}{Mc^2},
\eeq
yielding the gravitomagnetic clock effect obtained in \cite{Cohen_Mashhoon:1993}. It is an
interesting fact that (\ref{elektro.gem-clock}), as well as its magnetic counterpart
(\ref{elektro.mce-opp}), neither depends on the coupling constant nor on the distance $r$
of the test masses. The reason for this is seen by inspection of (\ref{elektro.deltapsip}) and
(\ref{elektro.periode1}): both the square of the Kepler frequency and the induced velocity change
depend on $G/r^3$ ($\Qb\rightarrow 2GM$ in (\ref{elektro.deltapsip})) and hence it will not appear
in the expression $T_+-T_-\propto\Delta\ppsi/\ppsik^2$. This result is a consequence of the $r^{-3}$
fall-off of the dipolar magnetic field. In a homogeneous magnetic field, for instance, $\Delta\ppsi$
will be independent of the distance and the time difference will thus become $T_+-T_-\propto r^3$.
Further, the absence of the gravitational constant $G$ in (\ref{elektro.gem-clock}) allows the
time difference to become relatively large; inserting the appropriate values for the Earth into
(\ref{elektro.gem-clock}) yields $T_+-T_-\sim 10^{-7}$ s.

\section{An alternative clock effect}
\label{sect4}
If we introduce the angle $\xi=\kom+\gom\cos i+\calm$, where $\kom,\gom,\calm$ and $i$ are the
argument of pericenter, the longitude of the ascending node, the mean anomaly and the orbital
inclination, respectively, its temporal rate of change is given by
\beq
\frac{d\xi}{dt}=\frac{d\kom}{dt}+\frac{d\gom}{dt}\cos i-\gom\sin
i\frac{di}{dt}+\frac{d\calm}{dt}.
\eeq
Upon inserting the secular contributions of the oblateness of the Earth as well as the
corresponding Schwarz\-schild and Lense-Thirring rates we find that the time needed for
$\xi$ to cover an angle of $2\pi$ is given by 
\beq\label{Txi}
T_\xi=\frac{2\pi}{\bar{n}}
 \left[1+J_2\frac{3\rerde^2\left(1+\sqrt{1-e^2}\right)}{4a^{2}(1-e^2)^2}\left(1-3\cos^2 i\right)-
 \frac{3\bar{n}^2a^2}{c^2(1-e^2)}+\frac{4\bar{n}J\cos i}{c^2M(1-e^2)^{3/2}}\right],
\eeq
where $a,\ e$ and $\bar{n}$ are the semimajor axis,
eccentricity and mean motion, and $\rerde,\,J$ and $M$ are the
radius, angular momentum and mass of the Earth, respectively;
$J_2$ is the lowest zonal harmonic coefficient in the multipolar
expansion of the terrestrial gravitational field.
For satellites along identical but opposite orbits, all but the last term in (\ref{Txi}) cancel
when taking the time difference in the periods of $\xi$, revealing the gravitomagnetic signature
\beq\label{diff-Txi}
T_\xi^+-T_\xi^-=\frac{16\pi J\cos i}{c^2M(1-e^2)^{3/2}},
\eeq
which could in principle be measured by means of two counter-revolving satellites around the Earth.
We note that (\ref{diff-Txi}) reduces to equation (14) of \cite{Iorio_Lichtenegger:2005} in the case
of nearly equatorial and circular orbits.

\section{Error analysis}
\label{sect5}
\begin{vchfigure}[b]
\centering
  \includegraphics[width=0.9\textwidth]{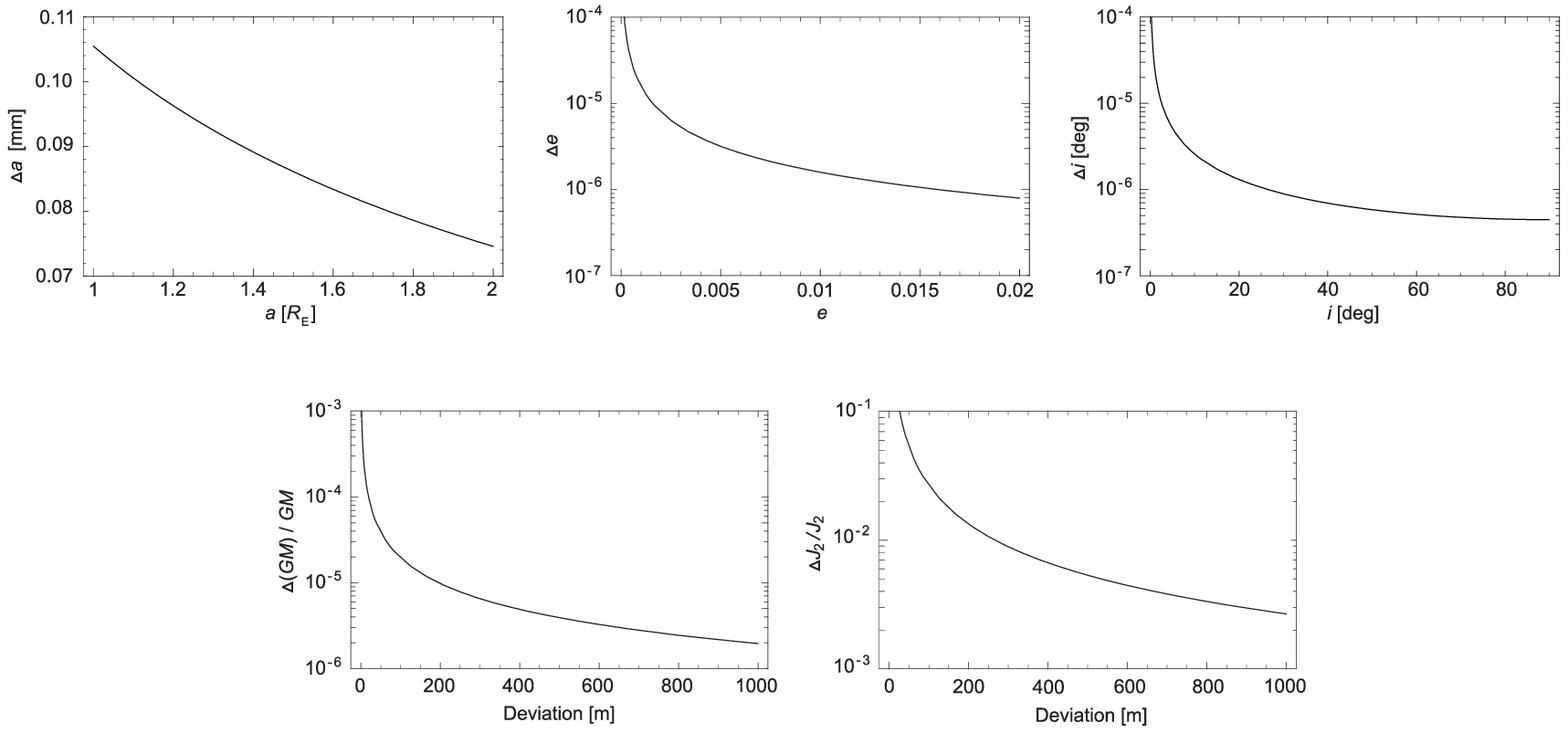}
\vchcaption{Maximum tolerable error of various values for a nominal orbit
of $a=12\,000$ km, $e=0.01$ and $i=63.4^\circ$.}
\label{fig:2}
\end{vchfigure}
Equation (\ref{diff-Txi}) holds only for the ideal case that the two
orbits are identical and the orbital elements are exactly known.
In a real experiment, both the orbits will differ from each other
due to injection errors and the elements are only determined with
limited accuracy. These residuals will contribute to the
difference in equation (\ref{diff-Txi}) and must hence be kept less
than the Lense-Thirring difference. Although equatorial orbits
would be favorable, however, since most satellite laser ranging
stations are located in the northern hemisphere, inclined orbits
are more suitable from an observational point of view.
For a nominal orbit with $a=12\,000$ km, $e=0.01$ and $i=63.4^\circ$,
which represents a frozen perigee configuration, 
the upper panel in Fig. 2 displays the maximum allowable uncertainty of
$\Delta a,\,\Delta e$ and $\Delta i,$ as a function of the corresponding
orbital elements, while the lower panel illustrates the maximum allowable
change of $\Delta(GM)/GM$ and $\Delta J_2/J_2$ with respect to the deviation
from the nominal orbit of $a=12\,000$ km.
As can be seen, a relative accuracy of $\Delta a/a<6\times
10^{-12},\ \Delta e/e< 1\times 10^{-4}, \,\Delta i/i<8\times 10^{-9}$ is required,
while the uncertainties in the values of $GM$ and $J_2$ are of minor importance,
since they are known to an accuracy of $\Delta(GM)/GM\sim 10^{-9}$ and
$\Delta J_2/J_2\sim 10^{-7}$ \cite{iers:2003}. With about 2500 revolutions per
year, after 4 years of observation and for a target error of 1\%, the semimajor
axis must be known with an accuracy of about 1 part in $10^{9}$, i.e. to a few cm,
$e$ to 1\% and $i$ to 0.0001\%. The stringent limit for $i$ is due to the high
inclination of the nominal orbit; for near equatorial orbits this limit becomes
less severe \cite{Iorio_Lichtenegger:2005}. Finally, we note that care must also
be taken in the accurate modelling of all non-gravitational perturbations. In this
case, however, one would greatly benefit from the experience gained in the
LAGEOS-LAGEOS II Lense-Thirring experiment.

\section{Conclusion}
It has been shown that the gravitomagnetic clock effect for spherical orbits has a simple
correspondence with the electromagnetic case. The difference in the sidereal periods of two
counter-revolving charges about a central spinning charge is due both to the different velocities
of the test charges and to the precession of the orbital plane. The slower prograde velocity, which
in the gravitational case is against the common notion of the dragging of inertial frames, can be
considered as a consequence of the Lenz rule. The gravitomagnetic clock effect is obtained by simply
substituting the electric and magnetic charges by the corresponding gravitoelectric and gravitomagnetic
ones. With regard to the gravitational perturbations, a key issue for a possible observation of the clock
effect with two counter-orbiting satellites is the stringent requirement in the accuracy of the inclinations.


\begin{thebibliography}{10}

\bibitem{Cohen_Mashhoon:1993}
J.M.~Cohen and B.~Mashhoon,
  Phys.\,Lett.\,A \textbf{181}, 353--358 (1993).

\bibitem{Mashhoon_et_al:1999}
B.~Mashhoon, F.~Gronwald, and D.S.~Theiss,
  Ann.\,Physik \textbf{8}, 135--152 (1999).

\bibitem{Mashhoon_et_al:2001a}
B.~Mashhoon, F. Gronwald, and H. Lichtenegger,
  in: Gyros, Clocks, Interferometers...: Testing Relativistic Gravity in
  Space, edited by C.\ L\"ammerzahl, C.\,W.\,F.\ Everitt and F.\,W.\
  Hehl (Springer, Berlin, 2001), pp. 83-108.

\bibitem{Mashhoon_et_al:2001b}
B. Mashhoon, L. Iorio, and H. Lichtenegger,
  Phys. Lett. A \textbf{292}, 49-57 (2001).

\bibitem{Iorio_et_al:2002}
L.~Iorio, H.~Lichtenegger, and B.~Mashhoon,
  Class. Quantum Grav. \textbf{19}, 39-49 (2002).

\bibitem{Everitt_et_al:2001}
C.W.F. Everitt et al.,
  in: Gyros, Clocks, Interferometers...: Testing Relativistic Gravity in
  Space, edited by C.\ L\"ammerzahl, C.\,W.\,F.\ Everitt and F.\,W.\
  Hehl (Springer, Berlin, 2001), pp. 52-82.

\bibitem{Schiff:1960}
L.~Schiff,
  Phys. Rev. Lett. \textbf{4}, 215--217 (1960).

\bibitem{www:stanford}
For information about Gravity Probe B, see http://einstein.stanford.edu.

\bibitem{Lense_Thirring:1918}
J.~Lense and  H.~Thirring,
  Phys. Z. \textbf{19}, 156-163 (1918).

\bibitem{Mashhoon_et_al:1984}
B.~Mashhoon, F. W. Hehl, and D. S. Theiss,
  Gen. Rel. Grav. \textbf{16}, 711--750 (1984).

\bibitem{Iorio_Morea:2004}
L.~Iorio and A.~Morea,
  Gen. Rel. Grav. \textbf{36}, 1321--1333 (2004).

\bibitem{Ciufolini_Pavlis:2004}
I. Ciufolini and E.C. Pavlis,
  Nature \textbf{431}, 958--960 (2004).

\bibitem{Iorio:2006}
L.~Iorio, J. of Geodesy \textbf{80}, 128--136 (2006).

\bibitem{Iorio:2005}
L.~Iorio,
  First preliminary evidence of the general relativistic gravitomagnetic field of the Sun and
  new constraints on a Yukawa-like fifth force, (2005) [http://arxiv.org/abs/gr-qc/0507041].

\bibitem{Pitjeva:2005}
E.V. Pitjeva,
  Astron. Lett. \textbf{31}, 340--349, 2005.

\bibitem{Mashhoon:1993}
B.~Mashhoon,
  Phys.\,Lett.\,A \textbf{173}, 347--354 (1993).

\bibitem{Iorio_Lichtenegger:2005}
L. Iorio and H.I.M. Lichtenegger,
  Class. Quantum Grav. \textbf{22}, 119--132 (2005).

\bibitem{iers:2003}
D.D. McCarthy and G. Petit (eds.), IERS Conventions (2003), IERS Technical Note No. 32
 (Verlag des Bundes\-amts f\"ur Kartographie und Geod\"asie, Frankfurt am Main, 2004).




\end{thebibliography}
\end{document}